\documentstyle[epsfig,longtable]{aipproc}

\begin{document}

\title{Thermal-viscous instabilities in the accretion disk of GRS~1915+105}
 
\author{T.~Belloni$^{1}$,
        M.~M\'endez$^{1,2}$,
        A.R.~King$^{3}$,
        M.~van der Klis$^{1}$,
        J.~van Paradijs$^{1,2}$
}

\address{
$^{1}$Astronomical Institute ``Anton Pannekoek'', University of
       Amsterdam and Center for High-Energy Astrophysics,
       Kruislaan 403, NL-1098 SJ Amsterdam, the Netherlands\\
$^{2}$Facultad de Ciencias Astron\'omicas y Geof\'{\i}sicas,
       Universidad Nacional de La Plata, Paseo del Bosque S/N,
       1900 La Plata, Argentina\\
$^{3}$Astronomy Group, University of Leicester,
       Leicester LE1 7RH, United Kingdom
}

\maketitle

\begin{abstract}

We use data obtained with the PCA onboard RXTE to analyze the spectral
variations of the black hole candidate and superluminal source
GRS~1915+105.  Our results suggest that despite the complicated structure
of the X-ray light curve, all spectral changes observed in this source
can be explained as a sudden disappearing of the inner part of the
accretion disk around the compact object, followed by a slower
re-filling of the emptied region.  
The duration of an event and the size of the
disappearing region fit remarkably well the expected radius dependence
of the viscous time scale for the radiation-pressure dominated region of
an accretion disk.

\end{abstract}

\section*{Introduction}

The X--ray source and black hole candidate (BHC) GRS~1915+105 was the
first galactic object to show relativistic jets with apparent
superluminal motions \cite{mr94}.  It is at a distance of 12.5~kpc
\cite{ch96}, and the jet is oriented at an angle of $70^\circ$ with
respect to the line of sight.  The Rossi X-ray Timing Explorer (RXTE)
has observed this source systematically since April 1996, and since then
GRS~1915+105 has shown a remarkable richness in variability:  quasi-periodic
burst-like events, deep regular dips and strong quasi-periodic
oscillations, and long quiescent periods
\cite{gmr96,mrg96,cst97,bmk97a,tcs97}.  Here we show that its
complicated light curve can be described by the rapid
appearance and disappearance of emission from the inner accretion disk.

\section*{Observations and Data Analysis}

\begin{figure}[ht] 
\centerline{\epsfig{file=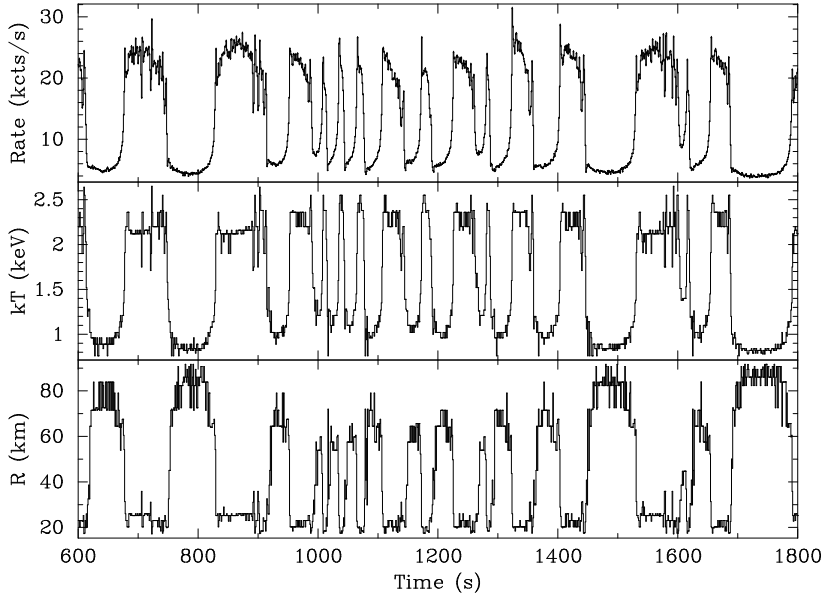,width=3.5in,angle=-90}}
\vspace{5pt}
\caption{Upper panel: 2.0--40.0 keV PCA light curve.
Time zero corresponds to 1997 June 18th 14:36 UT.
Middle and lower panels: corresponding inner radius
and temperature (see text).}
\label{fig1}
\end{figure}

RXTE observed GRS~1915+105 in many occasions.  Here we present the
results of the analysis of one of these observations, the one obtained
on 1997 June 18 starting at 14:36 UT and ending at 15:35 UT, as it
reproduces within one day most of the variability observed from this
source.  The upper panel of Figure \ref{fig1} shows 1200~s of the
2-40~keV light curve.  It consists of a sequence of `bursts' of
different duration with quiescent intervals in between.  All bursts
start with a well-defined sharp peak and decay faster than they rise.
The longer bursts show oscillation (or sub-bursts) towards the end.

To study the evolution of the spectrum of GRS~1915+105, we first
produced a color-color (C-C) diagram of the whole observation (see
\cite{bmk97b}, Figure 3) which we divided in small regions.  For all the
populated regions in this diagram we accumulated 1 second time
resolution energy spectra in 48 energy bands.  We measured a background
spectrum from a blank sky observation, which we normalized to the
highest energy channels where the contribution of the source was
negligible.  We subtracted this background spectrum from each of the
source energy spectra.  We used the latest detector response matrix
available, and we added a systematic error of 2\% to account for the
calibration uncertainties.  For each of the regions in the C-C diagram
we fitted the data with a ``standard'' spectral model for BHCs,
consisting of a disk-blackbody (DBB) model and a power law plus
interstellar absorption.  To avoid problems due to the background
subtraction, and as we were only interested in the properties of the DBB
component, we limited our fits to energies below 30~keV.  Since both the
distance to this system and the inclination of the accretion disk are
known (we assume that the jet is perpendicular to the disk), we can
derive the inner radius of the accretion disk directly from the fits.

\begin{figure}[ht] 
\centerline{\epsfig{file=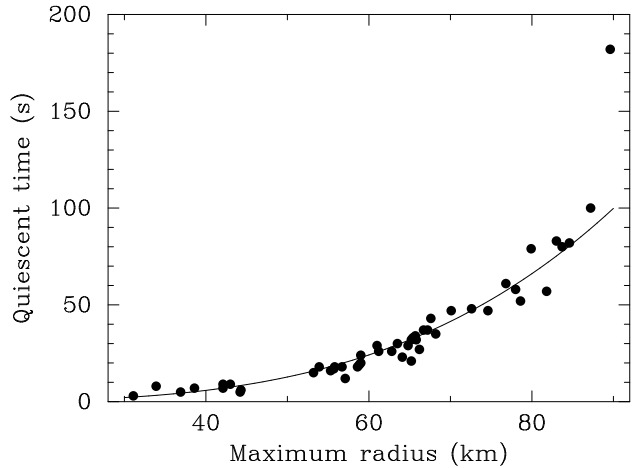,width=2.9in,angle=-90}}
\vspace{5pt}
\caption{Correlation between the total length of an event and maximum
inner radius of the disk. The line is the best fit with a
power law with fixed index $\gamma = 3.5$. The last point
has been excluded from the fit.}
\label{fig2}
\end{figure}

We used the time resolved energy spectra to study the variations of the
inner radius and the temperature of the disk as a function of time
(bottom two panels of Figure \ref{fig1}).  During bursts the temperature
is above 2~keV and the radius is stable around 20~km.  During quiescent
phases, the temperature drops to less than 1~keV and the radius increases.
There is a strong correlation between the length of an event and the
largest radius reached (Figure \ref{fig2}).

We produced similar C-C diagrams for a number of other RXTE observations
of GRS~1915+105.  All the observations that we analyzed can be fitted in
the same manner, except that of 1996 June 16th \cite{gmr96}.  All the
quasi-periodic bursts observed in many of the observations \cite{tcs97}
are consistent with repetitive short events like the ones described
here.

\section*{Discussion}

The previous results can be interpreted as follows:  The large amplitude
changes reflect the emptying and replenishing of the inner accretion
disk, caused by a viscous-thermal instability \cite{bmk97a}.  The small
radius observed during the quiescent period can be identified with the
innermost stable orbit around the black hole, while the large radius
during the burst phase represents the radius of the emptied section of
the disk.  The smaller oscillations are simply failed attempts to empty
the inner disk.  Figure \ref{fig1} shows that all variations, from major
events \cite{bmk97a} to small oscillations at the end of a large event,
can be modeled in exactly the same manner.

Both the spectral evolution and the duration of the event are determined
only by the radius of the missing inner section of the accretion disk.
For a large radius, the drop in flux is larger, and the time needed to
re-fill the empty part of the disk is longer.  It is natural to
associate the length of the quiescent part of an event to the viscous
time scale of the radiation-pressure dominated part of the accretion
disk \cite{bmk97a}.  This can be expressed as $t_{\rm visc} = 0.30
\alpha^{-1} M_1^{-1/2} R_7^{7/2}\dot M_{18}^{-2}$~seconds, where
$\alpha$ is the viscosity parameter, $R_7$ is the radius in units of
$10^7$ cm, $M_1$ the central object mass in solar masses, and $\dot
M_{18}$ is the accretion rate in units of $10^{18}$ g/s.  Notice that
even the largest radii derived here are well within the
radiation-pressure dominated part of the disk \cite{bmk97a}.  The line
in Figure 2 represents the best fit to the data with a relation of the
form $t_{\rm q} \propto R^{7/2}$.  The fit is excellent, with the
exception of the point corresponding to the longest event.

An event can therefore be pictured in the following way.  At the start
of a quiescent period, the disk has a central hole, whose radius is
R$_{max}$.  The hole is either empty or filled with gas whose radiation
is too soft to be detected.  Slowly the disk is re-filled by a steady
accretion rate $\dot M_0$ from outside.  Each annulus of the disk will
move along the lower branch of its S-curve in the $\dot M-\Sigma$ plane
trying to stabilize at $\dot M_0$ \cite{bmk97a}.  The surface gravity
increases as the annulus moves towards the unstable point at a speed
determined by the local viscous time scale.  During this period, no
changes are observed in the radius of the hole, since the matter inside
does not radiate in the PCA band.  The observed accretion rate is $\dot
M_0$.  At some point, one of the annuli will reach the unstable point
and switch to the high-$\dot M$ state, where the accretion rate is
larger than $\dot M_0$, causing a chain-reaction that will ``switch on''
the inner disk.  The observed accretion rate is now higher than the
external value $\dot M_0$.  A smaller, hot radius is now observed.  At
the end of the outburst, the inner disk runs out of fuel and switches
off, either jumping back to the $\dot M < \dot M_0$ state or emptying
completely.  A new hole is formed and a new cycle starts.  Notice that
in this scenario the more ``normal'' state for the source is the one at
high count rates, where the disk extends all the way to the innermost
stable orbit:  in this state the energy spectrum is similar to that of
conventional BHCs.

\end{document}